\documentclass[submission,copyright]{eptcs}
\providecommand{\event}{HSB 2012} 

\usepackage{graphicx} 
\usepackage{amsmath}
\usepackage{caption} 
\usepackage{array} 
\usepackage{tikz} 

\title{A Model of the Cellular Iron Homeostasis Network Using Semi-Formal Methods for Parameter Space Exploration}

\author{Nicolas Mobilia
\institute{UJF-Grenoble 1 / CNRS\\TIMC-IMAG UMR 5525,\\Grenoble, F-38041, France}
\email{nicolas.mobilia@imag.fr}
\and
Alexandre Donz{\'e}
\institute{EECS Department\\University of California Berkeley\\Berkeley, CA 94720 USA}
\email{donze@eecs.berkeley.com}
\and
Jean Marc Moulis
\institute{UJF-Grenoble 1 / CNRS UMR 4952,\\Institut de Recherches en Technologies et Sciences pour le Vivant,\\17 rue des Martyrs,\\38054 Grenoble}
\email{jean-marc.moulis@cea.fr}
\and
{\'E}ric Fanchon
\institute{UJF-Grenoble 1 / CNRS\\TIMC-IMAG UMR 5525,\\Grenoble, F-38041, France}
\email{eric.fanchon@imag.fr}
}

\def\titlerunning{A Model of the Cellular Iron Homeostasis Network}
\def\authorrunning{N. Mobilia, A. Donz{\'e}, J.M. Moulis, {\'E}. Fanchon}

\newcommand{\IRP}{\mathit{IRP}}
\newcommand{\Fe}{\mathit{Fe}}
\newcommand{\TfR}{\mathit{TfR1}}
\newcommand{\FPN}{\mathit{FPN1a}}
\newcommand{\Ft}{\mathit{Ft}}
\newcommand{\f}{\varphi}

\begin{document}

\maketitle

\begin{abstract}
  This paper presents a novel framework for the modeling of biological
  networks. It makes use of recent tools analyzing the robust
  satisfaction of properties of (hybrid) dynamical systems. The main
  challenge of this approach as applied to biological systems is to
  get access to the relevant parameter sets despite gaps in the
  available knowledge. An initial estimate of useful parameters was
  sought by formalizing the known behavior of the biological network
  in the STL logic using the tool Breach. Then, once a set of
  parameter values consistent with known biological properties was
  found, we tried to locally expand it into the largest possible valid
  region. We applied this methodology in an effort to model and better
  understand the complex network regulating iron homeostasis in
  mammalian cells. This system plays an important role in many
  biological functions, including erythropoiesis, resistance against
  infections, and proliferation of cancer cells.
\end{abstract}

\section{Introduction}

In cellular biology, the value of the parameters are uncertain (when a
measurement does exist) or not known at all (unmeasured or not
computable from first principles). It is well known that values of
kinetic constants obtained from in vitro measurements on purified
enzymes can be significantly different from the in vivo values, due to
interactions with other cell components, to regulatory modifications
of enzymes, to sequestration, to anomalous diffusion or still other
phenomena. In addition, the values available from the literature are
heterogeneous (measurement performed in different species, or on
different cell types, or in different conditions). Two sets of
measurements performed on identical cell types placed in supposedly
identical conditions can also be qualitatively different for multiple
reasons (undetected heterogeneity of cell populations, different
batches of antibodies used for detection, for instance). The fact that
the available data and pieces of information are generally too scarce
with respect to the size of dynamical cell models forces the modeler
to include every bit of available data, at the cost of increased
heterogeneity and thus increased uncertainty. But even with such an
inclusive strategy the amount of data remains inadequate to identify a
fully instantiated model.

The complexities and heterogeneity of both the parameter values and
the analytical expressions of the terms entering the differential
equations describing a biological system implies such a system results
from a family of dynamical systems, instead of a single one. The
properties which formalize experimentally observed behaviors should
thus be robust with respect to the uncertainties mentioned above.

These issues are considered in the present paper and the semi-formal
method implemented to explore the parameter space is applied to an
important question in biology: how iron homeostasis is maintained in
mammalian cells? Iron is the most abundant transition metal required
by most forms of life. The largest part needed by animals ends up in
red blood cells, but all cell types do use the
metal~\cite{andrews_2012}. Iron is handled by specific molecules in
animal physiology, so as to avoid the deleterious reactions the iron
ions catalyze, conversion of oxygen derivatives into reactive species
(the reactive oxygen species, ROS) prominent among
them~\cite{galaris_pantopoulos_2008}. The mandatory requirement for
iron and the negative impact the metal may have on cellular function
call for a tight regulation of its
use~\cite{domenico_mcvey_2008}. Deregulation translates into cellular
oxidative damage: this links iron homeostasis to the redox balance in
animal, and other, cells. These connections underlie scores of (human)
pathological situations in which these homeostatic systems (iron and
redox homeostasis) are perturbed.

Our approach rests on the following points: we deal explicitly with
the parameter space, and we do not focus on a single instantiation of
the model parameters. When the parameter values are uncertain due to
the fact that they have been measured in different cell types or in
different conditions, the data are formally represented by
inequalities. In other words, we do not assign the measured value
directly to the parameter but rather impose inequality constraints on
that parameter. If there is no measurement at all, bounds are chosen
to define a physiologically reasonable domain. Observed behavioral
properties can be formalized using a temporal logic formalism. Then
the behaviors of the parametrized system are explored by sampling
predefined volumes of parameter space (including initial conditions
and model parameters), and the instantiated models that satisfy the
temporal logic formula are identified.
As illustrated below this very basic scheme needs some elaboration to
be applied to the modeling of a complex, yet simplified, biological
phenomenon.

The biological network we have built for iron homeostasis is presented
in section~\ref{biological_network}. The system of differential
equations is introduced in section~\ref{ODE_modeling} and its
dynamical properties explored in the following section.

\section{Biological network}
\label{biological_network}

We are considering regulation of iron handling in cell types which
acquire iron by endocytosis of the transferrin-transferrin receptor
complex (most of them) and express an iron exporter in the form of
ferroportin. Proper supply of iron to animal cells is required to
avoid deleterious effects of too much or too little iron as such
imbalance jeopardizes the proper function of cells and may lead to
death. Regulation of iron supply occurs at two levels, a systemic one
and a local one organized around Iron Regulatory Proteins. The latter
are the main focus of this study since they directly act on the
molecular nodes of the considered network.

\subsection{Species}

\begin{figure}[!ht]
  \begin{minipage}{0.55\linewidth}
    \centering
    \includegraphics[width=\linewidth]{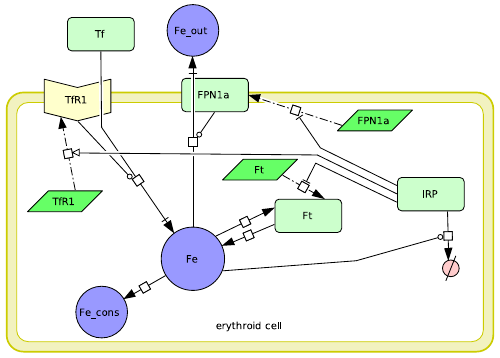}\\~\\~\\
    A)
  \end{minipage}
  \hfill
  \begin{minipage}{0.45\linewidth}
    \centering
    \tikzstyle{cercle}=[circle,thick,draw,minimum size=0.25cm]
    \begin{tikzpicture}[thick,auto,->,scale=0.8]
    
    \node[cercle] (Fe)    at (0,0)     {$\Fe$};
    \node[cercle] (TfR1)  at (1,4.5)   {$\TfR$}; 
    \node[cercle] (FPN1a) at (4.5,4.5) {$\FPN$}; 
    \node[cercle] (Ft)    at (3,1.5)   {$\Ft$};
    \node[cercle] (IRP)   at (6,0)     {$\IRP$};
    
    \draw (TfR1)  to [left,bend right=10]             node {+} (Fe);
    \draw (TfR1)  to [loop above]                     node {-} (TfR1);
    
    \draw (FPN1a) to [bend right=10,near end,above]   node {-} (Fe);
    \draw (FPN1a) to [loop above]                     node {-} (FPN1a);
    
    \draw (Fe)    to [bend right=45,below]            node {-} (IRP);
    \draw (Fe)    to [loop left,near end,above]       node {-} (Fe);
    
    \draw (Ft)    to [bend right=5,right,near end]    node {+} (Fe);
    \draw (Ft)    to [loop below]                     node {-} (Ft);
    
    \draw (IRP)   to [bend right=10,right]            node {-} (FPN1a);
    \draw (IRP)   to [bend right=15,above,near start] node {+} (TfR1);
    \draw (IRP)   to [below,bend right=5]             node {-} (Ft);
    \draw (IRP)   to [bend left=25,above,near start]  node {+} (Fe);
    \draw (IRP)   to [loop right,near start,above]    node {-} (IRP);

    \end{tikzpicture}\\
    B)
  \end{minipage}
  \caption{A) Schematic representation of the main biological processes
  involved in the cellular homeostatic control of iron.\\ The dashed
  arrows represent translation of mRNA into proteins. The arrow ending
  in an unfilled white arrow represents amplification of translation,
  while the lines with a broken ending arrowhead represent inhibition
  of translation. The lines ending with a combined perpendicular
  stroke and arrow represent iron transport through membranes. The
  arrow leading to a pink circle indicates degradation. Finally, the
  two regular arrows represent the loading/unloading of iron into/from
  the ferritins and the consumption of iron for the cell needs. The
  green rounded rectangles represent proteins, the green
  parallelograms represent mRNA, and the blue circles represent
  atoms. The yellow concave hexagon represents the transferrin
  receptor. This diagram was drawn with the software
  CellDesigner~\cite{funahashi_morohashi_2003}.\label{fig:ntwk_iron}\\
  B) Interaction graph defined by the equations. The signs on the
  arrows indicate whether the interaction is positive or
  negative.\label{fig:interaction_graph}}
\end{figure}

The considered network is shown in Figure~\ref{fig:ntwk_iron}~A. It
contains five species: Iron Regulatory Protein (IRP, no difference is
made between the two known proteins displaying this function), iron
(Fe), ferritin (Ft), the transferrin receptor (TfR1) and ferroportin
(FPN1a). Moreover, transferrin (Tf) is a protein which carries iron in
the blood stream throughout the body, and which can bind zero, one, or
two iron atoms. It is the main provider of iron to most cells and, as
such, it is an input of the network. The circle labeled by Fe\_out is
the iron exported out of the cell. The circle labeled by Fe\_cons
represents the iron consumed for the cell needs.

The IRP are proteins which can bind to a specific hairpin-like
structure on mRNA, called Iron Responsive Element (IRE). Depending on
the position of the IRE on the non-coding sequence of mRNA, two main
outputs may result:
\begin{itemize}
\item If the IRE is located at the beginning of the mRNA, in the 5'
  untranslated region (UTR), the binding of the IRP prevents the
  translation machinery from binding to the mRNA and translating it,
  so the binding of the IRP leads to a reduced quantity of the
  corresponding protein;
\item If the IRE is located at the end of the mRNA (3'-UTR), the
  binding of the IRP inhibits the degradation of this mRNA, leading to
  an increase in the quantity of translated protein.
\end{itemize}
In the presence of enough iron, the regulatory activity of the IRP is
decreased by different molecular
mechanisms~\cite{wang_pantopoulos_2011}.

The iron species (Fe) represents the cellular iron which is available
for cellular processes. The actual form of this iron in biological
cells is still a topic of discussion. A virtual species, which we call
``Fe'', is introduced in order to model the usable iron in the
cell. If this amount of intracellular iron becomes too low, the cell
cannot maintain vital functions and dies.

Ferritin is a protein which stores
iron~\cite{arosio_levi_2010}. The complex mechanisms of iron
loading into and release from ferritin are not considered
here. The mRNA of both ferritin subunits contain an IRE on their
5'-UTR regions, and the IRP inhibit their translation.

Transferrin receptors are located on the surface of the
cell~\cite{aisen_2004}. The fixation of transferrin on its receptor
leads to the endocytosis of the complex. This process involves several
steps and releases iron into the cell. Transferrin receptor mRNA
contain five IRE in the 3'-UTR region, which is stabilized upon IRP
binding. The density of transferrin receptor at the cell surface is
thus correlated with the IRP activity.

Finally, ferroportin (FPN1a) is an iron exporter located at the
surface of some cells~\cite{mayr_janecke_2010}. A form of FPN1a
contains an IRE on its 5'-UTR mRNA region, so the translation is
inhibited by the presence of IRP.

\subsection{Behavior}
\label{sec:behavior}

A qualitative description of the system has been obtained through a
large body of biological experiments and we selectively summarize them
here.

If the amount of iron is sufficient (\emph{iron-replete} situation),
the cell is in a stationary state. Assuming that the iron provision
stays constant, the concentration of each protein described in
Figure~\ref{fig:ntwk_iron}~A does not vary. In this state, the IRP
activity is expected to be relatively low. In the case of too much
iron (which is not studied here), the saturation level of transferrin
increases, the IRP activity reaches a minimum and inhibits translation
of the iron importer (TfR1) to minimize cellular iron input.

From the iron-replete stationary state, if the
cells become \emph{iron-depleted}, the inhibition of IRP activity by
iron decreases. Thus, the increasing IRP activity leads to an increase
in the transferrin receptor concentration and to a decrease in
ferroportin concentration. This, in turn, leads to an augmentation in
the amount of iron entering the cell and a reduction in the amount of
iron leaving the cell. Moreover, the iron loaded in ferritin is
released to convert stored iron into a form available for biosynthetic
purposes. This release allows the cell to face a temporary reduction
in the level of iron (typically because of a reduction in the iron
input) for a certain amount of time. When ferritin is depleted, and if
there is still no iron input mediated by the receptor TfR1, the cell
runs out of iron and enters in a cellular death process. The IRP
regulating system is effective as long as the iron level remains
low. When the iron level regains an acceptable level again, the IRP
activity decreases, and the amount of filled ferritin goes up,
regenerating the stock of stored iron.

\section{ODE modeling}
\label{ODE_modeling}

In this section we describe the differential equations that we derived
to specifically describe the evolution of the concentration of each
species over time. Some processes in the iron homeostasis network are
in fact a composition of many reactions, and the choice of an
analytical expression to represent them is not straightforward. This
is the case of the iron entry mediated by TfR1, which involves
endocytosis, and of the loading of iron in ferritin, for example.

%
%
%

\subsection{Iron equation}
\label{def_iron}

In the cell, when the level of iron becomes low, some iron-proteins
release their iron content, by protein degradation for instance, which
contributes to refill the iron pool; however, only iron release from
ferritin is considered in our model without affecting the general
validity of the reasoning. The equation defining the iron
concentration is:
\begin{equation}
\mathit{\frac{\mathrm{d}Fe}{\mathrm{d}t} = k_{Fe\_input}\cdot{}TfR1\cdot{}Tf_{sat} - n_{Ft}\cdot\frac{\mathrm{d}Ft}{\mathrm{d}t} - k_{Fe\_export}\cdot{}Fe\cdot{}FPN1a - k_{Fe\_cons}\cdot{}Fe}
\end{equation}

The import of iron into the cell is proportional to the concentration
of transferrin receptor and to the average amount of iron bound to
transferrin (called transferrin saturation): $\mathit{Tf_{sat}}$. The
second term describes the storage or release of iron due to the
synthesis or degradation of ferritin. This variation is equal to the
number of iron atoms per ferritin molecule (aka $\mathit{n_{Ft}}$)
times the variation in the ferritin concentration (we consider that
(i) iron release occurs only when ferritin is degraded (ii)
synthesized ferritins are very quicky filled with iron). The third
term describes the export of iron out of the cell as proportional to
the iron and FPN1a concentration. The export parameter is
$\mathit{k_{Fe\_export}}$. Finally, within the cell, iron is used for
many purposes; this consumption of iron is represented by the last
term in our equation ($\mathit{k_{Fe\_cons}\cdot{}Fe}$). For the sake
of simplicity, we consider this consumption term (which represents a
whole set of cellular processes) to be proportional to the iron
concentration. The export of iron by FPN1 forms that are not regulated
by IRP can be accounted for by this term.

\subsection{IRP equation}
\label{def_IRP}

The equation describing the IRP concentration is:
\begin{equation}
\mathit{\frac{\mathrm{d}IRP}{\mathrm{d}t} = k_{IRP\_prod} - k_{Fe{\rightarrow}IRP}\cdot{}sig^+\left(Fe,\theta_{Fe{\rightarrow}IRP}\right)\cdot{}IRP - k_{IRP\_deg}\cdot{}IRP}
\end{equation}

We define a constant rate of production of IRP by
$\mathit{k_{IRP\_prod}}$. The concentration of iron regulates the
activity of IRP. The degradation of IRP is described by a constant
basal term ($\mathit{k_{IRP\_deg}\cdot{}IRP}$) and an iron-related
term
($\mathit{k_{Fe{\rightarrow}IRP}\cdot{}sig^+\left(Fe,\theta_{Fe{\rightarrow}IRP}\right)\cdot{}IRP}$). Then,
if the iron level is significantly below the threshold
$\mathit{\theta_{Fe{\rightarrow}IRP}}$, the degradation rate is
$\mathit{k_{IRP\_deg}\cdot{}IRP}$. If the iron concentration is
significantly above this threshold, the degradation rate is
$\mathit{(k_{Fe{\rightarrow}IRP}+k_{IRP\_deg})\cdot{}IRP}$, where
$\mathit{k_{Fe{\rightarrow}IRP}}$ is the parameter describing the
inhibition of IRP by iron.

The sigmoid function is defined such that:
\begin{equation*}
sig^+(x,\theta) = \frac{x^n}{x^n+\theta^n}
\end{equation*}
where $x$ is a concentration variable, $\theta$ is the threshold and
$n$ defines the steepness of the sigmoid. In our model, we use $n=30$,
which corresponds to a very steep sigmoid.

\subsection{Ferritin equation}
\label{def_ferritin}

To model the temporal evolution of ferritin, we assume that each
ferritin protein is rapidly filled with iron atoms, right after its
synthesis is completed. In other words the amount of empty ferritin is
negligible.
The variable Ft thus represents the concentration of filled ferritin
proteins within the cell. The equation describing the evolution of the
concentration of ferritin is:
\begin{equation}
\mathit{\frac{\mathrm{d}Ft}{\mathrm{d}t} = k_{Ft\_prod} - k_{IRP{\rightarrow}Ft}\cdot{}sig^+\left(IRP,\theta_{IRP{\rightarrow}Ft}\right) - k_{Ft\_deg}\cdot{}Ft}
\end{equation}
We consider a basal production rate described by
$\mathit{k_{Ft\_prod}}$. The second term describes the regulation by
IRP. We use a sigmoidal regulation here, but we could consider other
dynamics. If the activity of IRP is above the threshold
$\mathit{\theta_{IRP{\rightarrow}Ft}}$, the production rate is lowered
by $\mathit{k_{IRP{\rightarrow}Ft}}$. It follows that
$\mathit{k_{IRP{\rightarrow}Ft}}$ has to be lower or equal to
$\mathit{k_{Ft\_prod}}$: as IRP inhibits translation, this inhibition
should never lead to a negative production rate. Finally, the third
term ($\mathit{k_{Ft\_deg}\cdot{}Ft}$) describes the spontaneous
degradation of ferritin.

\subsection{Ferroportin equation}
\label{def_ferroportin}

The equation describing the ferroportin concentration is very similar
to that of ferritin. The equation is:
\begin{equation}
\mathit{\frac{\mathrm{d}FPN1a}{\mathrm{d}t} = k_{FPN1a\_prod} - k_{IRP{\rightarrow}FPN1a}\cdot{}sig^+(IRP,\theta_{IRP{\rightarrow}FPN1a}) - k_{FPN1a\_deg}\cdot{}FPN1a}
\end{equation}

The first term describes the basal production of FPN1a. The second
term expresses the regulation of IRP on the translation of FPN1a. With
the same reasoning as previously, the regulation parameter
$\mathit{k_{IRP{\rightarrow}FPN1a}}$ has to be lower or equal to the
production parameter $\mathit{k_{FPN1a\_prod}}$. The dynamics of IRP
regulation on Ft and FPN1a are considered similar because both FPN1a
and Ft mRNAs have a single IRE located in the 5'-UTR region without
experimental evidence of major differences between the two.

\subsection{Transferrin receptor equation}
\label{def_transferrin_receptor}

The concentration of the transferrin receptor evolves according to the
following equation:
\begin{equation}
\mathit{\frac{\mathrm{d}TfR1}{\mathrm{d}t} = k_{TfR1\_prod} + k_{IRP{\rightarrow}TfR1}\cdot{}IRP - k_{TfR1\_deg}\cdot{}TfR1}
\end{equation}

The production of TfR1 includes a basal rate $\mathit{k_{TfR1\_prod}}$
and it is increased proportionally to the IRP concentration. This is
taken into account in the term
$\mathit{k_{IRP{\rightarrow}TfR1}\cdot{}IRP}$. This term is different
from those describing the regulation of ferritin and ferroportin in an
effort to consider the binding of several IRPs to a single TfR mRNA
molecule (since it contains several IREs). The half-life of TfR1 mRNA
is considered proportional to the IRP concentration, although this is
a rough approximation~\cite{erlitzki_long_2002}. The last term of the
equation represents the spontaneous degradation.

\subsection{Interaction graph}
\label{interaction_graph}

The interaction graph is a useful tool to visualize the pattern of
interactions in a system of differential equations. It is defined from
the Jacobian matrix of the system. Each non-zero element in this
matrix indicates that a species influences another species. The nodes
of the interaction graph are the dynamical variables of the system
(concentration of biological species), and each arc represents the
influence of a node on another, i.e. a non-zero element of the
Jacobian matrix. The sign of this element determines whether the
influence is positive or negative. Of course it is possible that the
sign of the elements depends on the point of phase space at which the
Jacobian is evaluated, but in our model the signs are constant. The
interaction graph we obtain is shown in
Figure~\ref{fig:interaction_graph}~B.









We define a negative (resp. positive) circuit (closed path) as a
succession of arrows in the interaction graph such that the product of
the signs labelling these arrows is negative (resp. positive). This
graph contains three negative circuits of length greater than one
(IRP$\rightarrow$Fe$\rightarrow$IRP~;
IRP$\rightarrow$FPN1a$\rightarrow$Fe$\rightarrow$IRP and
IRP$\rightarrow$TfR1$\rightarrow$Fe$\rightarrow$IRP), and five
negative loops corresponding to spontaneous degradation or
consumption. It also contains one positive circuit
(~IRP$\rightarrow$Ft$\rightarrow$Fe$\rightarrow$IRP~). The presence of
these circuits, together with the presence of non-linear terms in the
equations, generate a complex dynamical behavior.

\section{Methodology}

In this section we present our strategy to study the system. We first
translate all available biological data into inequalities on
parameters and variables. Then, we formally define the expected
behavior for the two modes described in Section~\ref{sec:behavior}:
the \emph{iron-replete} mode and the \emph{iron-depleted}
mode. Finally we look for a subspace which robustly verifies this
behavior.

\subsection{Parameter search space}
\label{initial_parameter_space}

The parameter space of the system is twenty-dimensional. To define an
initial parameter search space, we combined published data from
different sources as shown in table~\ref{tab:data}. Whenever possible,
the available data obtained with erythroleukemic cells have been used,
as these cells are relatively immature and proliferate, as considered
for the initial steady state.
\begin{table}[!ht]
\centering
\begin{tabular}{|m{0.46\linewidth}|m{0.46\linewidth}|}
\hline
In mouse macrophages, half-life of FPN1a seems to be more than 20 hours~\cite{knutson_oukka_2005} & $0 < \mathit{k_{FPN1a\_deg}}~\le~9.6\times10^{-6}~s^{-1}$\\
\hline
The production rate of TfR1 was estimated at 7$\times10^{-6}$ pg/cell/min as in~\cite{omholt_kefang_1998} & $1.0\times10^{-13} \le \mathit{k_{TfR1\_prod}} \le 2.0\times10^{-13}$ mol/L/s\\
\hline
The half-life of IRP is 12-15 hours in H1299 human lung cancer cells~\cite{dycke_charbonnier_2008} & $1.28\times10^{-5} \le \mathit{k_{IRP\_deg}} \le 1.6\times10^{-5}~\text{s}^{-1}$\\
\hline
In human erythroleukemia cell line K562, half-life of TfR1 is 8 hours~\cite{weissman_klausner_1986} & $2.0\times10^{-5} \le \mathit{k_{TfR1\_deg}} \le 3.0\times10^{-5}~\text{s}^{-1}$\\
\hline
The TfR1 production rate enhanced by IRP is 6$\times10^{-5}$ pg/cell/min~\cite{omholt_kefang_1998} & $4.2\times10^{-5} \le \mathit{k_{IRP{\rightarrow}TfR1}} \le 14.4\times10^{-5}~\text{s}^{-1}$\\
\hline
In human erythroleukemia cell line K562, iron input is up to $7.2{\pm}2.4\times10^4$ Fe atoms/cell/min~\cite{shvartsman_fibach_2010} & $2\times10^{-2} \le \mathit{k_{Fe\_input}} \le 3.9\times10^{-2}~\text{s}^{-1}$\\
\hline
In mammalian cells, the number of iron atoms in ferritin up to 4500~\cite{arosio_levi_2010} & $0 < \mathit{n_{Ft}} \le 4500$\\
\hline
Human normal saturation level of transferrin is within 25\% and 45\%~\cite{moirand_mortaji_1997}, we take 30\% & $\mathit{Tf_{sat}} = 0,3$ (in iron-replete situation)\\
\hline
In human erythroid cells, a fraction of the iron exporter (FPN1) is not IRP-regulated~\cite{cianetti_segnalini_2005}. The amount of iron needed for biosynthesis (``consumed'') is higher than the exported one. & $\mathit{k_{Fe\_cons}} \ge \mathit{k_{Fe\_export}\cdot{}FPN1a}$ (in iron-replete situation)\\
\hline
A higher limit for the iron pool in erythroleukemic cells is set at 2 $\mu$mol/L~\cite{epsztejn_kakhon_1997} & $\Fe \le 2\times10^{-6}$ mol/L\\
\hline
In rats liver, the IRP concentration in around 0.11 pmol/mg of cytosolic protein under normal conditions~\cite{chen_blemings_1998} & $3\times10^{-9} \le \IRP \le 10.7\times10^{-9}$ mol/L (in iron-replete situation)\\
\hline
In human erythroleukemic (K562) cells, there are 140,000 transferrin receptors by cell~\cite{hunt_marshall_1986} & $1.0\times10^{-8} \le \TfR \le 10.0\times10^{-8}$ mol/L (in iron-replete situation)\\ 
\hline
\end{tabular}
\caption{The left column shows the data collected in the literature;
  the right column describes the corresponding interval on parameters
  and variables, and the inequalities deduced.\label{tab:data}}
\end{table}

We then obtained intervals for eight parameters, including one in the
iron-replete situation. Moreover, we derived an inequality linking
$\mathit{k_{Fe\_cons}}$, $\mathit{k_{Fe\_export}}$ and
$\FPN$. Finally, we found data on the value of three variables ($\Fe$,
$\IRP$ and $\TfR$) among which two are related to the iron-replete
situation. 

In order to evaluate the search space for the twelve other
parameters, we proceeded by analogy. For example, we had no
information on the basal production rate of ferritin, ferroportin and
IRP, but we knew an interval for the production rate of the transferrin
receptor. We then assumed that the basal production rate of
ferroportin was neither more than 1000 times higher than the highest
possible transferrin receptor basal production rate, nor 1000 times
lower than the lowest. Concerning the thresholds parameters, we simply
assumed that their possible intervals were the same as for the
corresponding protein concentrations.




\subsection{The iron-replete steady state}
\label{steady_state}

The iron-replete situation is modeled by a steady state, so we do not
consider evolutions of the variables over time. In this state, the
value of the variable representing the iron concentration is expected
to be significantly higher than the threshold
$\mathit{\theta_{Fe{\rightarrow}IRP}}$. We can thus infer that the
sigmoidal term
$\mathit{sig^+\left(Fe,\theta_{Fe{\rightarrow}IRP}\right)}$ is very
close to one, and we replace it. The value of the variable
representing the IRP activity is also expected to be largely below the
thresholds $\mathit{\theta_{IRP{\rightarrow}Ft}}$ and
$\mathit{\theta_{IRP{\rightarrow}FPN1a}}$. We then consider that the
two sigmoidal functions
$\mathit{sig^+\left(IRP,\theta_{IRP{\rightarrow}Ft}\right)}$ and
$\mathit{sig^+(IRP,\theta_{IRP{\rightarrow}FPN1a})}$ are very close to
zero. These simplifications contribute to reducing the complexity of
the system. Note that this amounts to approximating the sigmoid by
step functions, and consequently the system could be viewed as an
hybrid system.

We developed two complementary approaches for studying the system in
the iron-replete steady state. The first consists in a mathematical
analysis of the system. The second consists in performing deductions
from the quantitative data.
%

Looking for a steady state, we set the derivatives in the model to
zero and solved the resulting algebraic system using the symbolic
solver Maxima~\cite{maxima}. This way, we could prove the existence of
one unique stationary point, given by Equations~(\ref{eqn_Fe_stable})
to~(\ref{eqn_TfR1_stable}).
\begin{figure}[!ht]
\begin{align}
Fe = {} & \frac{Tf_{sat}\cdot k_{FPN1a\_deg}\cdot k_{Fe\_input}\cdot \left( k_{TfR1\_prod}\cdot \left( k_{IRP\_deg}+k_{Fe{\rightarrow}IRP}\right)+k_{IRP{\rightarrow}TfR1}\cdot k_{IRP\_prod}\right) }{\left( k_{Fe\_export}\cdot k_{FPN1a\_prod}+k_{Fe\_cons}\cdot k_{FPN1a\_deg}\right) \cdot \left( k_{IRP\_deg}+k_{Fe{\rightarrow}IRP}\right) \cdot k_{TfR1\_deg}}\label{eqn_Fe_stable}\\
Ft = {} & \frac{k_{Ft\_prod}}{k_{Ft\_deg}}\label{eqn_Ft_stable}\\
FPN1a = {} & \frac{k_{FPN1a\_prod}}{k_{FPN1a\_deg}}\label{eqn_FPN1a_stable}\\
IRP = {} & \frac{k_{IRP\_prod}}{k_{IRP\_deg}+k_{Fe{\rightarrow}IRP}}\label{eqn_IRP_stable}\\
TfR1 = {} & \frac{\left( k_{IRP\_deg}+k_{Fe{\rightarrow}IRP}\right)\cdot k_{TfR1\_prod}+k_{IRP{\rightarrow}TfR1}\cdot k_{IRP\_prod}}{\left(k_{IRP\_deg}+k_{Fe{\rightarrow}IRP}\right) \cdot k_{TfR1\_deg}} \label{eqn_TfR1_stable}
\end{align}
\end{figure}

We then considered the stability of this stationary point by computing
its Jacobian matrix since an unstable stationary point cannot
obviously represent an observed cellular state. The obtained Jacobian
matrix is shown in equation~(\ref{eqn:jacobian_sig}).
\begin{equation}
\centering
\begin{pmatrix}
\mathit{-FPN1a\cdot{}k_{Fe\_export}-k_{Fe\_cons}} & \mathit{Tf_{sat}\cdot{}k_{Fe\_input}} & \mathit{-Fe\cdot{}k_{Fe\_export}} & \mathit{k_{Ft\_deg}\cdot{}n_{Ft}} & 0\cr
0 & \mathit{-k_{TfR1\_deg}} & 0 & 0 & \mathit{k_{IRP{\rightarrow}TfR1}}\cr
0 & 0 & \mathit{-k_{FPN1a\_deg}} & 0 & 0\cr
0 & 0 & 0 & \mathit{-k_{Ft\_deg}} & 0\cr
0 & 0 & 0 & 0 & \mathit{-k_{IRP\_deg}-k_{Fe{\rightarrow}IRP}}
\end{pmatrix}
\label{eqn:jacobian_sig}
\end{equation}
As this matrix is upper triangular, its eigenvalues are equal to its 
diagonal elements. It follows that all eigenvalues are real and
negative, which proves the stability of the stationary point.

Another way to study the steady state consists in using the specific
data we have on the variables at the stationary state and propagate
them on the parameter space. For this purpose, we use
Realpaver~\cite{granvilliers_benhamou_2006}, an interval solver. The
solver uses the algebraic equations between parameters and variables,
as well as defined intervals and constraints, as inputs for the
calculation. The aim of Realpaver is to reduce the provided intervals
and consequently reduce the parameter space. The results are the
following:
\begin{itemize}
\item \ \hbox to 6cm{$\mathit{k_{FPN1a\_deg}} \ge 1.0\times10^{-13}~\text{s}^{-1}$\hfill} (initial interval: [$0.0,~9.6\times10^{-6}$]\,)
\item \ \hbox to 6cm{$\mathit{k_{FPN1a\_prod}} \le 9.6\times10^{-11}$ mol/L/s\hfill} (initial interval: [$1.0\times10^{-18},~1.0\times10^{-10}$]\,)
\item \ \hbox to 6cm{$\mathit{k_{IRP\_prod}} \ge 3.84\times10^{-14}$ mol/L/s\hfill} (initial interval: [$1.0\times10^{-18},~1.0\times10^{-10}$]\,)
\item \ \hbox to 6cm{$\mathit{k_{Fe\_cons}} \le 3.39\times10^{-4}~\text{s}^{-1}$\hfill} (initial interval: [$0.0,~1.0$]\,)
\item \ \hbox to 6cm{$\mathit{k_{Fe{\rightarrow}IRP}} \le 3.33\times10^{-2}~\text{s}^{-1}$\hfill} (initial interval: [$1.0\times10^{-9},~1.0$]\,)
\end{itemize}

Realpaver returns intervals which are reduced by many orders of
magnitude for four of the parameters. These results prove that some
subspaces of the parameter space do not contain valid parameter sets,
and then, need not be considered when searching for an parameter set
producing an adequate iron-replete steady state. This will reduce the
search of the parameter space in the next step (analysis of the
dynamical behavior).

Besides, Realpaver makes no difference between variables and
parameters considered in the algebraic system, they are all variables
(unknowns). Consequently it also propagates the constraints from the
parameters to the variables, and provides deductions on three
variables in the iron-replete situation:
\begin{itemize}
\item \ \hbox to 6cm{$\FPN \ge 1.04\times10^{-13}$ mol/L\hfill} (initial interval: [$1.0\times10^{-13},~1.0\times10^{-5}$]\,)
\item \ \hbox to 6cm{$\Fe \ge 3.5\times10^{-8}$ mol/L\hfill} (initial interval: [$0.0,~2.0\times10^{-6}$]\,)
\item \ \hbox to 6cm{$\TfR \le 8.7\times10^{-8}$ mol/L\hfill} (initial interval: [$1.0\times10^{-8},~1.0\times10^{-7}$]\,)
\end{itemize} 

These results, provided by Realpaver in less than one second on a
Core~2~Duo 3~GHz with 8~Gb of RAM, can be considered as deductions
made from the initial data. Indeed, the difficulty in finding a valid
parameter set is that steady state values must be within their
intervals. These two approaches facilitate the manual search of an
initial parameter space. The analytical solution of the equations
indicates, for each parameter, the variables impacted by a change in
the parameter value. We use this information to define an order for
checking variables, such that there are parameters allowing us to
adjust a variable value but not changing the value of the already
correct variables. Consequently, to find an initial valid parameter
set, one has only to tune some parameters to fix the value of the
first variable, then tune some other parameters to fix the value of
the second variable, and so on. If the value of one variable cannot be
set in its interval, one has to step back and change the value of
parameters used to fix the value of the previous variable. The
variable order we use is: IRP, TfR1 (through the adjustment of
$\mathit{k_{IRP{\rightarrow}TfR1}}$, $\mathit{k_{TfR1\_prod}}$ and
$\mathit{k_{TfR1\_deg}}$), FPN1a, Fe (through $\mathit{k_{Fe\_cons}}$,
$\mathit{k_{Fe\_export}}$ and $\mathit{k_{Fe\_input}}$) and finally
Ft. Using this method, we found the following initial parameter set,
denoted as $\mathcal{P}_0$:\\
\begin{tabular}{rcll cc rcll}
$k_{\TfR\_prod}$&=&$1.7\times10^{-13}$ &mol/L/s                &&&$k_{\FPN\_deg}$&=&$5.0\times10^{-6}$ &s$^{-1}$\\
$k_{\TfR\_deg}$&=&$2.4\times10^{-5}$ &s$^{-1}$                 &&&$k_{\Fe\_input}$&=&$3.0\times10^{-2}$ &s$^{-1}$\\
$n_{\Ft}$&=&$400$ &-                                           &&&$k_{\Ft\_prod}$&=&$7\times10^{-11}$ &mol/L/s\\
$k_{\IRP\_prod}$&=&$8.0\times10^{-12}$ &mol/L/s                &&&$k_{\Ft\_deg}$&=&$5.0\times10^{-3}$ &s$^{-1}$\\
$k_{\Fe\_export}$&=&$300$ &L/mol/s                             &&&$k_{\Fe{\rightarrow}\IRP}$&=&$1.0\times10^{-3}$ &s$^{-1}$\\
$\theta_{\IRP{\rightarrow}\FPN}$&=&$3.0\times10^{-8}$ &mol/L   &&&$\theta_{\Fe{\rightarrow}\IRP}$&=&$1.5\times10^{-7}$ &mol/L\\
$k_{\IRP{\rightarrow}\FPN}$&=&$5.0\times10^{-13}$ &mol/L/s     &&&$k_{\IRP\_deg}$&=&$1.4\times10^{-5}$ &s$^{-1}$\\
$k_{\IRP{\rightarrow}\TfR}$&=&$1.4\times10^{-4}$ &s$^{-1}$     &&&$k_{\FPN\_prod}$&=&$2.5\times10^{-12}$ &mol/L/s\\
$k_{\Fe\_cons}$&=&$3.0\times10^{-4}$ &s$^{-1}$                 &&&$\theta_{\IRP{\rightarrow}\Ft}$&=&$3.0\times10^{-8}$ &mol/L\\
$k_{\IRP{\rightarrow}\Ft}$&=&$7\times10^{-11}$ &mol/L/s
\end{tabular}

It will be used as the initial parameter set for the exploration of
the parameter space, as explained in the following section.

\subsection{Dynamics in iron-depleted situation}

We are now interested in the behavior of the network when the input of
iron is cut. As a consequence, we need to compute the evolution of the
system over time. For that, we use
Breach~\cite{donze_2010,donze_fanchon_2011}, a tool based on
Matlab/C++. This tool allows not only to simulate differential
systems, but also to express temporal logic formula and to check
whether a trajectory satisfies or not a given formula.

We introduce briefly STL (Signal Temporal Logic). A more complete
description can be found in~\cite{donze_fanchon_2011}. Initially
created for verifying the correctness of reactive programs, temporal
logics provide languages to express properties over time-dependent
phenomena for any kind of dynamical system. STL and its quantitative
semantics \cite{DonzeM10} specializes into systems with continuous
time and with real-valued states. An STL formula has the following
syntax:
\begin{equation}
\f =  \mu\verb? | (?\f_1\verb?) and (?\f_2\verb?) | ev_[a,b] (?\f\verb?) | alw_[a,b] (?\f\verb?)?
\end{equation}
where $\mu$ is an inequality constraint between expressions involving
variables (written \verb?var[t]?), constants or derivatives of a
variable over time (written \verb?ddt{var}[t]?). A formula is
evaluated at each requested time (thereafter called ``the current
time''). A formula with a \verb?and? statement is true if both $\f_1$
and $\f_2$ are true at the current time. The \verb?ev_[a,b]? statement
is true if $\f$ holds at least once between the current time plus
\verb?a? and the current time plus \verb?b?. If the interval is
omitted, then \verb+a+ is 0 and \verb+b+ is $\inf$, that is, the
formula is true if $\f$ holds at least once, anytime after the current
time. The statement \verb?alw_[a,b]? is true if $\f$ holds between
current time plus \verb?a? and current time plus \verb?b?. Again, if
the interval is omitted then \verb+a+ is 0 and \verb+b+ is $\inf$
meaning in that case that the formula must hold all the time.

The experiment we want to simulate is the cut-off of iron entry for
cells initially in the iron-replete situation. The total duration of
the simulation is 48 hours, as the qualitative description of the
iron-depleted situation tells us that significant changes can be seen
in 24 hours. First we want the simulations to stabilize in a steady
state corresponding to the iron-replete situation. Then, at time t=6
hours, we cut the entry of iron by setting $\mathit{Tf_{sat}}$ to zero. We
expect the system to turn on the IRP regulatory system in order to
maintain a correct iron level for at least ten hours. To identify
parameter values for which the simulations satisfy all theses
properties, we use temporal logic formula.

We first want the system to be in a steady state corresponding to the
iron-replete situation. To avoid an inappropriately long stabilization
phase, we specify initial conditions close to the value of the
stationary state. To ensure that the system reaches a steady state, we
use the following temporal logic formula:
\begin{equation}
\begin{aligned}
\f_{S1} = \verb|abs(ddt{Fe}[t]    / Fe[t])    < 1.0e-4|\\
\f_{S2} = \verb|abs(ddt{Ft}[t]    / Ft[t])    < 1.0e-4|\\
\f_{S3} = \verb|abs(ddt{FPN1a}[t] / FPN1a[t]) < 1.0e-4|\\
\f_{S4} = \verb|abs(ddt{IRP}[t]   / IRP[t])   < 1.0e-4|\\
\f_{S5} = \verb|abs(ddt{TfR1}[t]  / TfR1[t])  < 1.0e-4|
\end{aligned}
\label{eq:s1}
\end{equation}
(To make the reading easier, all formula dealing with the iron-replete
steady state will begin by ``$\f_S$'', formula related to parameters
will begin by ``$\f_P$'' and formula related to the behavior will
begin by ``$\f_B$''.) To check that the value of each variable at the
steady state is in the intervals shown in table~\ref{tab:data}, or
deduced in~\ref{steady_state}, we use the following formula:
\begin{equation}
\begin{aligned}
\f_{S6} = \verb|(3.5e-8 < Fe[t])                             |\\
\f_{S7} = \verb|(3.0e-9 < IRP[t]) and (IRP[t] < 1.07e-8)     |\\
\f_{S8} = \verb|(1.0e-8 < TfR1[t]) and (TfR1[t] < 8.7e-8)    |\\
\f_{S9} = \verb|(1.04e-13 < FPN1a[t]) and (FPN1a[t] < 1.0e-5)|\\
\f_{S10} = \verb|(1.0e-13 < Ft[t]) and (Ft[t] < 1.0e-5)      |
\end{aligned}
\label{eq:s2}
\end{equation}
As the maximal concentration of iron is $2\times10^{-6}$ mol/L all the
time, and not only at the steady state, we do not enforce this
property here.

Then, when the iron input is switched off, and the iron level becomes
too low, the ferritins release iron. This release is due to a lower
ferritin production rate, which provokes a decrease of ferritin
concentration. This can only be triggered by crossing the threshold
$\mathit{\theta_{IRP{\rightarrow}Ft}}$. This agrees with the previous
statement that the variable describing IRP activity is lower than
$\mathit{\theta_{IRP{\rightarrow}Ft}}$ under iron-replete
conditions. We verify that property with the following formula:
\begin{equation}
\f_{S11} = \verb|IRP[t] < theta_IRP_Ft|\label{eq:s11}
\end{equation}
This increase of IRP activity can only be triggered by a decrease in
iron concentration important enough to cross the threshold
$\mathit{\theta_{Fe{\rightarrow}IRP}}$. This is in line with the
already indicated inequality
$\mathit{Fe~>~\theta_{Fe{\rightarrow}IRP}}$ under iron-replete
conditions. The corresponding formula is:
\begin{equation}
\f_{S12} = \verb|Fe[t] > theta_Fe_IRP|\label{eq:s12}
\end{equation}
Moreover, when the iron level becomes too low, the FPN1a concentration
is expected to become lower. This, also, can only be triggered by a
crossing by IRP of the threshold
$\mathit{\theta_{IRP{\rightarrow}FPN1a}}$. That means that in the
iron-replete steady state the FPN1a concentration is above this
threshold, which is expressed by this formula:
\begin{equation}
\f_{S13} = \verb|IRP[t] < theta_IRP_FPN1a|\label{eq:s13}
\end{equation}
Nevertheless , as explained in section~\ref{def_ferroportin}
(resp. section~\ref{def_ferritin}), the regulation by IRP on FPN1a
($\mathit{k_{IRP{\rightarrow}FPN1a}}$) (resp. on Ft
($\mathit{k_{IRP{\rightarrow}Ft}}$)) cannot be higher that the basal
production rate of FPN1a ($\mathit{k_{FPN1a\_prod}}$) (resp. Ft
($\mathit{k_{Ft\_prod}}$)). The formula imposing that are:
\begin{equation}
\f_{P1} = \verb|k_IRP_FPN1a <= k_FPN1a_prod|\label{eq:p1}
\end{equation}
and
\begin{equation}
\f_{P2} = \verb|k_IRP_Ft <= k_Ft_prod| \label{eq:p2}
\end{equation}

The activation of the regulatory system leads to the conservation of
the iron level. We express that by enforcing the existence of a
plateau of this level after shutting the iron import off. We want this
plateau to exist for at least ten hours. To avoid incorrect
simulations for which the concentration of this plateau is zero, we
set it at, at least, one tenth of the concentration in the
iron-replete situation. The following property expresses that:
\begin{equation}
\begin{aligned}
\f_{B1} = \verb|ev_[6*3600, inf] (alw_[0, 10*3600]                       |\\
         \verb|          ( (|\f_{S1}\verb|)  and  (Fe[t] > 0.01*Fe[4*3600]) )) |\\
\end{aligned}
\label{eq:b1}
\end{equation}

When the iron stored in ferritin is exhausted, the iron concentration
eventually decreases to zero. Ferritin disappearance precedes iron
depletion which is not rescued by replenishment from any internal
store, in contrast to the transient phase immediately following
removal of iron input from transferrin
(Figure~\ref{fig:simu_sig}~A. Otherwise, it would mean that the cell
has stored iron but is not using it, which contradicts the conditions
set in~\ref{def_iron}. This implies that the parameter describing the
strength of the regulation by the IRP
(i.e. $\mathit{k_{IRP{\rightarrow}Ft}}$) is slightly lower or equal to
the production rate (i.e. $\mathit{k_{Ft\_prod}}$). We input this
constraint with the following formula:
\begin{equation}
\f_{P3} = \verb|k_Ft_prod*0.95 <= k_IRP_Ft| \label{eq:p3}
\end{equation}

Next, in table~\ref{tab:data}, two properties are expressed: the first
enforces that the iron concentration is never higher than
$2\times10^{-6}$ mol/L, and the second enforces that the inequality
$\mathit{k_{Fe\_cons}~\ge~k_{Fe\_export}\cdot{}FPN1a}$ holds in
iron-replete situation. The corresponding formula are:
\begin{equation}
\f_{B2} = \verb|alw (Fe[t] < 2e-6)|\label{eq:b2}
\end{equation}
and
\begin{equation}
\f_{S14} = \verb|k_Fe_cons > k_Fe_export*FPN1a[t]|\label{eq:s14}
\end{equation}

Finally, we have to define formula linking all the previous
formula. For this purpose, we first define the formula $\f_{Sall}$
which aggregates all formula related to the steady state. As we want
to be sure that, before the iron input is switched off, the system
reaches and stays at the steady state, we enforce that the properties
related to the steady state hold at least for one hour before the
sixth hour. This formula expresses that:
\begin{equation}
\f_{Sall} = \verb|ev_[0,6*3600] (alw_[0,3600] (|\f_{S1}\verb| and (|\f_{S2}\verb| and (|\f_{S3}\verb| and ( |\dots\verb| and |\f_{S14}\verb|))))|\label{eq:sall}
\end{equation}
We also define the formula $\f_{BPall}$ enforcing that all formula
related to parameters or behavior are satisfied:
\begin{equation}
\f_{BPall} = \f_{P1}\verb| and (|\f_{P2}\verb| and (|\f_{P3}\verb| and (|\f_{B1}\verb| and |\f_{B2}\verb|)))|\label{bpall}
\end{equation}
And finally, the formula $\f_{all}$, which enforces that all
properties are satisfied:
\begin{equation}
\f_{all} = \verb|(|\f_{Sall}\verb|) and (|\f_{BPall}\verb|)|\label{all}
\end{equation}

It is noticeable that no information on TfR1 in the iron-depleted
situation nor on the parameters acting on it appears in this
analysis. This is because, as we totally cut the entry of iron into
the cell, variations in the concentration of TfR1 do not change the
amount of incoming iron.

\subsection{A robust valid region}

\begin{figure}[!ht]
  \begin{minipage}{0.47\linewidth}
    \centering
    \includegraphics[width=\linewidth]{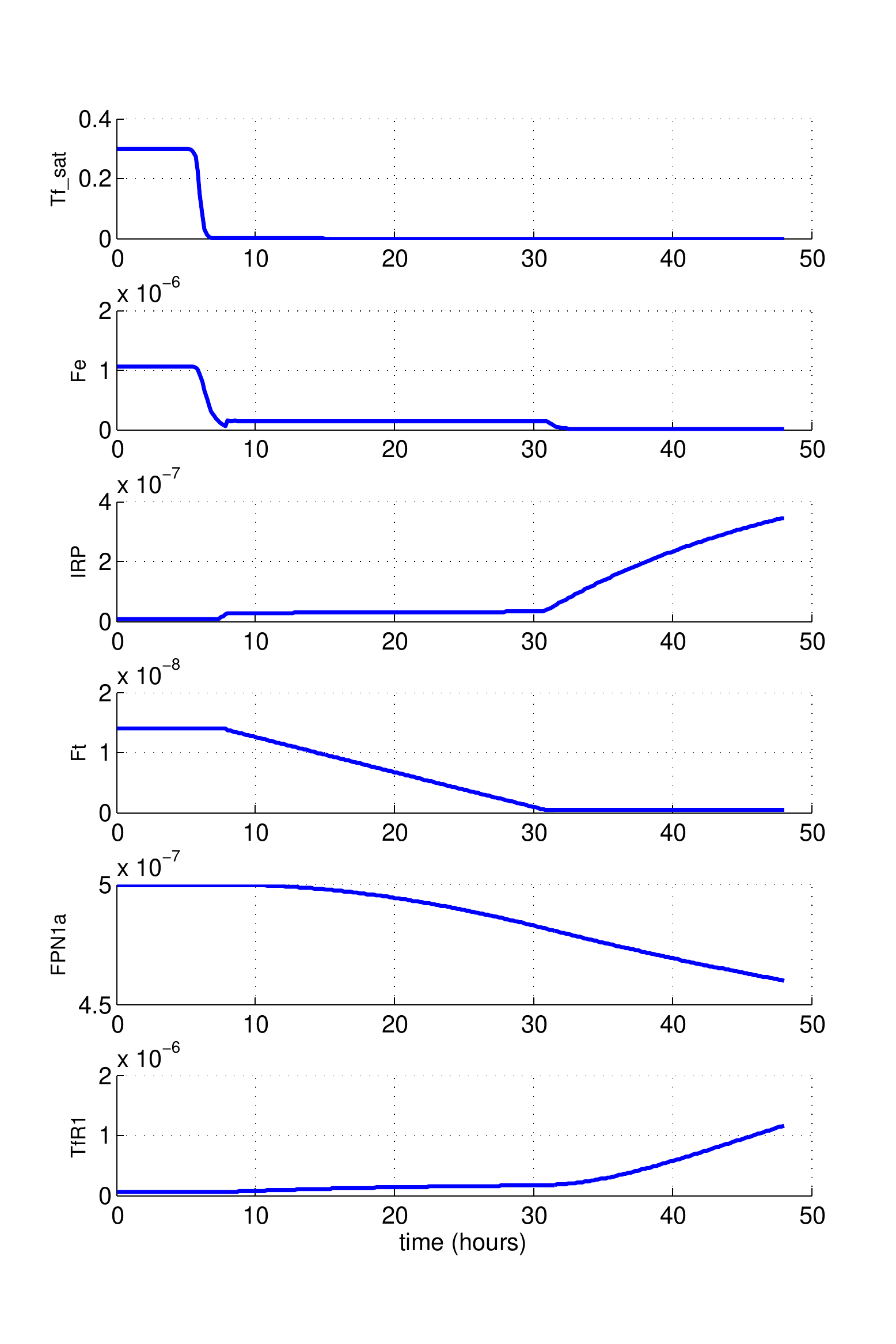}\\
    A)
  \end{minipage}
  \hfill
  \begin{minipage}{0.47\linewidth}
    \centering
    \includegraphics[width=\linewidth]{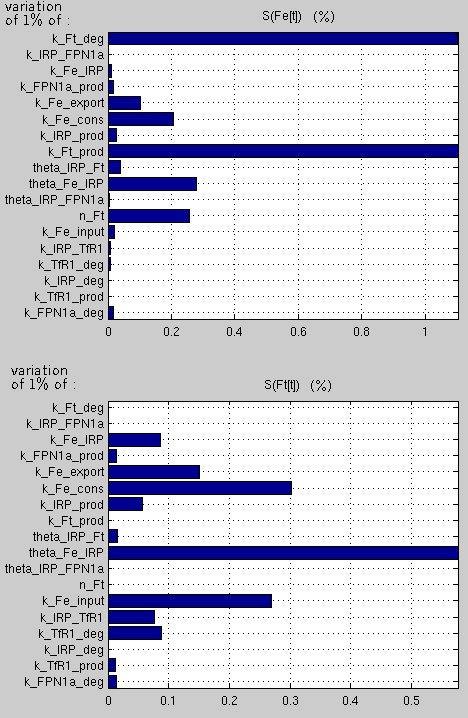}\\
    B)
  \end{minipage}
\caption{A) Characteristic simulation of the expected qualitative
  behavior. It shows that the system switches to a new mode
    when iron is lacking. The parameter set used is $\mathcal{P}_0$.
  At time t=6 hours, $\mathit{Tf_{sat}}$ is set to zero. The iron
  concentration consequently decreases, leading to an increase of IRP
  activity. This increase leads to a decrease in Ft concentration,
  keeping the iron level constant. This also leads to a decrease in
  FPN1a concentration and to an increase in TfR1
  concentration. During this phase when iron is kept constant,
    the concentration of iron is close to the threshold
    $\mathit{\theta_{Fe{\rightarrow}IRP}}$ and IRP activity is close
    to the thresholds $\mathit{\theta_{IRP{\rightarrow}FPN1a}}$ and
    $\mathit{\theta_{IRP{\rightarrow}Ft}}$. Once Ft concentration is
  zero, the iron concentration tends also to zero.\\
  B) Histogram showing the sensitivity of iron and ferritin
  concentration to variations of parameter value. These results are
  normalized: for example, a variation of 1\% in
  $\mathit{k_{Fe\_cons}}$ value leads to a variation of 0.2\% in Fe
  value. To avoid incorrect sensitivity due do an initial transient
  phase of large amplitude, we do not consider the first three hours
  when computing the sensitivity. Moreover, as the behavior of the
  cell (i.e. death) changes when iron is totally exhausted, we
  consider the sensitivity only between t=3 hours and t=20
  hours.\label{fig:simu_sig}}
\end{figure}

Having defined the formula which expresses the relevant properties, we
search for a subspace where all parameter sets verify these
properties, by first finding a valid instantiation of the parameters
and then exploring a neighborhood of this instantiation. In
section~\ref{steady_state}, we found an initial parameter set,
$\mathcal{P}_0$, satisfying constraints on the data for the steady
state. It turns out that the simulation computed with this parameter
set, shown in Figure~\ref{fig:simu_sig}~A, also verifies
$\f_{all}$. As the temporal formula describes the expected qualitative
behavior, this simulation is representative of the behavior of all
trajectories in the region that we are looking for. We look around
this point (in parameter space) to find a region exhibiting the
expected behavior. To ensure that the formula $\f_{P2}$ and $\f_{P3}$
are satisfied, we enforce that $\mathit{k_{IRP{\rightarrow}Ft}}$ is
computed equal to $0.97 \times \mathit{k_{ft\_prod}}$. Moreover, if
the parameter $\mathit{k_{FPN1a\_prod}}$ is lower than
$\mathit{k_{IRP{\rightarrow}FPN1a}}$, we switch them to force that the
formula $\f_{P1}$ is satisfied. So, the dimension of the search space
is slightly reduced.

To guide the search, we compute the sensitivity of the system to
parameters. We show in Figure~\ref{fig:simu_sig}~B the sensitivity of
iron and ferritin concentration to a variation of parameters
value. Sensitivity to ferroportin, transferrin receptor and IRP are
also computed, but are not shown. We define the sensitivity to a
parameter as the maximal sensitivity of the five variables of this
parameter. This sensitivity is used in the following manner: if the
system is very sensitive to a parameter, we will explore a close
neighborhood of this parameter. On the opposite, if the system
sensitivity to a parameter is low, we will explore a large
neighborhood. Nevertheless, it has to be noted that the system can be
very sensitive to a parameter, and that even with large variations on
this parameter the qualitative property is still satisfied. This
happens when the sensitivity which is detected affects an aspect of
the trajectories which is not captured by the property of interest.

To verify that all simulations corresponding to a parameter subspace
exhibit the expected behavior, we performed intensive trajectory
computations. We randomly picked 10000 parameter sets within this
subspace and checked the truth value of the formula $\f_{all}$. By
expanding as much as possible the parameter interval, we found an
hyper-rectangle centered on $\mathcal{P}_0$ such that all parameter
sets in this region exhibit the expected behavior. Some parameters of
$\mathcal{P}_0$ could change up to 70\% of their initial value, with
an average of 34\% for all, while conserving the behavior of the
system.

We observe that for the whole region of parameter space characterized
in the iron-depleted situation, the FPN1a concentration remains at a
rather high level, implying the considered cell keeps its ability to
export iron despite shortage of the metal.

\section{Conclusion}

Two models of the core iron homeostasis system have been published
previously. Omholt et al~\cite{omholt_kefang_1998} aim at building a
``unifying meta-model'', that is a model which is not tied to a
particular cell type. They include explicitly the two different Iron
Regulatory Proteins, IRP1 and IRP2. Both proteins have the same mRNA
targets, so they seem to be redundant, but they respond to the iron
level in the cell through different molecular mechanisms. The model of
Omholt et al~\cite{omholt_kefang_1998} does not explicitly include
ferritin nor ferroportin. In addition, all interactions in the network
are switch-like (sigmoidal kinetics). A numerical value was assigned
to all parameters except the sigmoid steepness, and the stability of
the unique steady state was monitored as the sigmoid steepness is
varied. The more recent paper of Chifman et
al~\cite{chifman_kniss_2012} is closer to our approach. This model
considers the same five biological species, but the analytical
expression of their differential system is different from ours. The
major differences are for the IRP, ferritin and ferroportin equations
(note that the graph shown in figure~1 of~\cite{chifman_kniss_2012}
does not follow the formal definition given here for an interaction
graph). The mathematical analysis of the parametrized model
of~\cite{chifman_kniss_2012} is general in the sense that it does not
rely on parameter instantiation. They showed that the model produces a
single steady state, the stability of which was studied by performing
simulations with sampled parameter values.

Here we have chosen an approach intermediate between Omholt et
al~\cite{omholt_kefang_1998} and Chifman et
al~\cite{chifman_kniss_2012}: a parametrized model has been built
without imposing numerical values to the parameters. The available
experimental data allowed us to apply inequality constraints between
the parameters, together with behavioral constraints expressed in a
temporal logic formalism. We could thus study not only steady states,
but also dynamical responses to perturbations, as, for instance, the
sharp decrease of iron input, starting from an iron-replete cell
state. The region of parameter space for which the model exhibits the
observed behavior has been characterized. It appeared that, in the
iron-depleted situation, the FPN1a concentration remains at a rather
high level, implying the considered cell kept its ability to export
iron despite shortage of the metal. This is a significant outcome of
modeling which justifies that another regulatory mechanism than that
of the IRP may be needed under these conditions. This additional
regulation may be provided by FPN1a degradation by hepcidin.

We plan to extend the present model. For example the influence of iron
on the inhibition of IRP activity involves the iron cluster
biosynthesis pathway which will be introduced in the model. Also, the
regulatory mechanisms of IRP could be described more precisely by
integrating mRNA concentrations in the model.
These extensions will be performed in a controlled and stepwise
manner. The qualitative behavior will be compared with that of the
core model presented here. This will give an insight into the
robustness of the dynamical property of interest.

\section*{Acknowledgments}
This work was supported by Microsoft Research through its PhD
Scholarship Programme.\\
The authors thank the R{\'e}gion Rh{\^o}ne-Alpes for financial support
within the Cible 2010 Program.

\bibliographystyle{eptcs}
\bibliography{biblio}

\begin{thebibliography}{10}
\providecommand{\bibitemdeclare}[2]{}
\providecommand{\surnamestart}{}
\providecommand{\surnameend}{}
\providecommand{\urlprefix}{Available at }
\providecommand{\url}[1]{\texttt{#1}}
\providecommand{\href}[2]{\texttt{#2}}
\providecommand{\urlalt}[2]{\href{#1}{#2}}
\providecommand{\doi}[1]{doi:\urlalt{http://dx.doi.org/#1}{#1}}
\providecommand{\bibinfo}[2]{#2}

\bibitemdeclare{article}{aisen_2004}
\bibitem{aisen_2004}
\bibinfo{author}{P.~\surnamestart Aisen\surnameend} (\bibinfo{year}{2004}):
  \emph{\bibinfo{title}{Transferrin receptor 1}}.
\newblock {\sl \bibinfo{journal}{The International Journal of Biochemistry \&;
  Cell Biology}} \bibinfo{volume}{36}(\bibinfo{number}{11}), pp.
  \bibinfo{pages}{2137--2143}, \doi{10.1016/j.biocel.2004.02.007}.

\bibitemdeclare{article}{andrews_2012}
\bibitem{andrews_2012}
\bibinfo{author}{N.~C. \surnamestart Andrews\surnameend}
  (\bibinfo{year}{2008}): \emph{\bibinfo{title}{{F}orging a field: the golden
  age of iron biology}}.
\newblock {\sl \bibinfo{journal}{Blood}}
  \bibinfo{volume}{112}(\bibinfo{number}{2}), pp. \bibinfo{pages}{219--230},
  \doi{10.1182/blood-2007-12-077388}.

\bibitemdeclare{article}{arosio_levi_2010}
\bibitem{arosio_levi_2010}
\bibinfo{author}{P.~\surnamestart Arosio\surnameend} \&
  \bibinfo{author}{S.~\surnamestart Levi\surnameend} (\bibinfo{year}{2010}):
  \emph{\bibinfo{title}{{C}ytosolic and mitochondrial ferritins in the
  regulation of cellular iron homeostasis and oxidative damage}}.
\newblock {\sl \bibinfo{journal}{Biochim. Biophys. Acta}}
  \bibinfo{volume}{1800}(\bibinfo{number}{8}), pp. \bibinfo{pages}{783--792},
  \doi{10.1016/j.bbagen.2010.02.005}.

\bibitemdeclare{article}{chen_blemings_1998}
\bibitem{chen_blemings_1998}
\bibinfo{author}{O.~S. \surnamestart Chen\surnameend}, \bibinfo{author}{K.~P.
  \surnamestart Blemings\surnameend}, \bibinfo{author}{K.~L. \surnamestart
  Schalinske\surnameend} \& \bibinfo{author}{R.~S. \surnamestart
  Eisenstein\surnameend} (\bibinfo{year}{1998}):
  \emph{\bibinfo{title}{{D}ietary iron intake rapidly influences iron
  regulatory proteins, ferritin subunits and mitochondrial aconitase in rat
  liver}}.
\newblock {\sl \bibinfo{journal}{J. Nutr.}}
  \bibinfo{volume}{128}(\bibinfo{number}{3}), pp. \bibinfo{pages}{525--535}.

\bibitemdeclare{article}{chifman_kniss_2012}
\bibitem{chifman_kniss_2012}
\bibinfo{author}{J.~\surnamestart Chifman\surnameend},
  \bibinfo{author}{A.~\surnamestart Kniss\surnameend},
  \bibinfo{author}{P.~\surnamestart Neupane\surnameend},
  \bibinfo{author}{I.~\surnamestart Williams\surnameend},
  \bibinfo{author}{B.~\surnamestart Leung\surnameend},
  \bibinfo{author}{Z.~\surnamestart Deng\surnameend},
  \bibinfo{author}{P.~\surnamestart Mendes\surnameend},
  \bibinfo{author}{V.~\surnamestart Hower\surnameend}, \bibinfo{author}{F.~M.
  \surnamestart Torti\surnameend}, \bibinfo{author}{S.~A. \surnamestart
  Akman\surnameend}, \bibinfo{author}{S.~V. \surnamestart Torti\surnameend} \&
  \bibinfo{author}{R.~\surnamestart Laubenbacher\surnameend}
  (\bibinfo{year}{2012}): \emph{\bibinfo{title}{{T}he core control system of
  intracellular iron homeostasis: a mathematical model}}.
\newblock {\sl \bibinfo{journal}{J. Theor. Biol.}} \bibinfo{volume}{300}, pp.
  \bibinfo{pages}{91--99}, \doi{10.1016/j.jtbi.2012.01.024}.

\bibitemdeclare{article}{cianetti_segnalini_2005}
\bibitem{cianetti_segnalini_2005}
\bibinfo{author}{L.~\surnamestart Cianetti\surnameend},
  \bibinfo{author}{P.~\surnamestart Segnalini\surnameend},
  \bibinfo{author}{A.~\surnamestart Calzolari\surnameend},
  \bibinfo{author}{O.~\surnamestart Morsilli\surnameend},
  \bibinfo{author}{F.~\surnamestart Felicetti\surnameend},
  \bibinfo{author}{C.~\surnamestart Ramoni\surnameend},
  \bibinfo{author}{M.~\surnamestart Gabbianelli\surnameend},
  \bibinfo{author}{U.~\surnamestart Testa\surnameend} \& \bibinfo{author}{N.~M.
  \surnamestart Sposi\surnameend} (\bibinfo{year}{2005}):
  \emph{\bibinfo{title}{{E}xpression of alternative transcripts of
  ferroportin-1 during human erythroid differentiation}}.
\newblock {\sl \bibinfo{journal}{Haematologica}}
  \bibinfo{volume}{90}(\bibinfo{number}{12}), pp. \bibinfo{pages}{1595--1606}.

\bibitemdeclare{article}{domenico_mcvey_2008}
\bibitem{domenico_mcvey_2008}
\bibinfo{author}{I.~\surnamestart De~Domenico\surnameend},
  \bibinfo{author}{D.~\surnamestart McVey~Ward\surnameend} \&
  \bibinfo{author}{J.~\surnamestart Kaplan\surnameend} (\bibinfo{year}{2008}):
  \emph{\bibinfo{title}{{R}egulation of iron acquisition and storage:
  consequences for iron-linked disorders}}.
\newblock {\sl \bibinfo{journal}{Nat. Rev. Mol. Cell Biol.}}
  \bibinfo{volume}{9}(\bibinfo{number}{1}), pp. \bibinfo{pages}{72--81},
  \doi{10.1038/nrm2295}.

\bibitemdeclare{inproceedings}{donze_2010}
\bibitem{donze_2010}
\bibinfo{author}{A.~\surnamestart Donz{\'e}\surnameend} (\bibinfo{year}{2010}):
  \emph{\bibinfo{title}{Breach, A Toolbox for Verification and Parameter
  Synthesis of Hybrid Systems}}.
\newblock In: {\sl \bibinfo{booktitle}{CAV}}, pp. \bibinfo{pages}{167--170},
  \doi{10.1007/978-3-642-14295-6\_17}.

\bibitemdeclare{article}{donze_fanchon_2011}
\bibitem{donze_fanchon_2011}
\bibinfo{author}{A.~\surnamestart Donz{\'e}\surnameend},
  \bibinfo{author}{{\'E}.~\surnamestart Fanchon\surnameend},
  \bibinfo{author}{L.~M. \surnamestart Gattepaille\surnameend},
  \bibinfo{author}{O.~\surnamestart Maler\surnameend} \&
  \bibinfo{author}{P.~\surnamestart Tracqui\surnameend} (\bibinfo{year}{2011}):
  \emph{\bibinfo{title}{Robustness Analysis and Behavior Discrimination in
  Enzymatic Reaction Networks}}.
\newblock {\sl \bibinfo{journal}{PLoS ONE}}
  \bibinfo{volume}{6}(\bibinfo{number}{9}), p. \bibinfo{pages}{e24246},
  \doi{10.1371/journal.pone.0024246}.

\bibitemdeclare{inproceedings}{DonzeM10}
\bibitem{DonzeM10}
\bibinfo{author}{A.~\surnamestart Donz{\'e}\surnameend} \&
  \bibinfo{author}{O.~\surnamestart Maler\surnameend} (\bibinfo{year}{2010}):
  \emph{\bibinfo{title}{Robust Satisfaction of Temporal Logic over Real-Valued
  Signals}}.
\newblock In: {\sl \bibinfo{booktitle}{FORMATS}}, pp. \bibinfo{pages}{92--106}.
\newblock \urlprefix\url{http://dx.doi.org/10.1007/978-3-642-15297-9_9}.

\bibitemdeclare{article}{dycke_charbonnier_2008}
\bibitem{dycke_charbonnier_2008}
\bibinfo{author}{C.~\surnamestart Dycke\surnameend},
  \bibinfo{author}{P.~\surnamestart Charbonnier\surnameend},
  \bibinfo{author}{K.~\surnamestart Pantopoulos\surnameend} \&
  \bibinfo{author}{J.~M. \surnamestart Moulis\surnameend}
  (\bibinfo{year}{2008}): \emph{\bibinfo{title}{{A} role for lysosomes in the
  turnover of human iron regulatory protein 2}}.
\newblock {\sl \bibinfo{journal}{Int. J. Biochem. Cell Biol.}}
  \bibinfo{volume}{40}(\bibinfo{number}{12}), pp. \bibinfo{pages}{2826--2832},
  \doi{10.1016/j.biocel.2008.05.015}.

\bibitemdeclare{article}{epsztejn_kakhon_1997}
\bibitem{epsztejn_kakhon_1997}
\bibinfo{author}{S.~\surnamestart Epsztejn\surnameend},
  \bibinfo{author}{O.~\surnamestart Kakhlon\surnameend},
  \bibinfo{author}{H.~\surnamestart Glickstein\surnameend},
  \bibinfo{author}{W.~\surnamestart Breuer\surnameend} \&
  \bibinfo{author}{I.~\surnamestart Cabantchik\surnameend}
  (\bibinfo{year}{1997}): \emph{\bibinfo{title}{{F}luorescence analysis of the
  labile iron pool of mammalian cells}}.
\newblock {\sl \bibinfo{journal}{Anal. Biochem.}}
  \bibinfo{volume}{248}(\bibinfo{number}{1}), pp. \bibinfo{pages}{31--40}.

\bibitemdeclare{article}{erlitzki_long_2002}
\bibitem{erlitzki_long_2002}
\bibinfo{author}{R.~\surnamestart Erlitzki\surnameend}, \bibinfo{author}{J.~C.
  \surnamestart Long\surnameend} \& \bibinfo{author}{E.~C. \surnamestart
  Theil\surnameend} (\bibinfo{year}{2002}): \emph{\bibinfo{title}{{M}ultiple,
  conserved iron-responsive elements in the 3'-untranslated region of
  transferrin receptor m{R}{N}{A} enhance binding of iron regulatory protein
  2}}.
\newblock {\sl \bibinfo{journal}{J. Biol. Chem.}}
  \bibinfo{volume}{277}(\bibinfo{number}{45}), pp.
  \bibinfo{pages}{42579--42587}, \doi{10.1074/jbc.M207918200}.

\bibitemdeclare{article}{funahashi_morohashi_2003}
\bibitem{funahashi_morohashi_2003}
\bibinfo{author}{.~\surnamestart Funahashi\surnameend},
  \bibinfo{author}{M.~\surnamestart Morohashi\surnameend},
  \bibinfo{author}{H.~\surnamestart Kitano\surnameend} \&
  \bibinfo{author}{N.~\surnamestart Tanimura\surnameend}
  (\bibinfo{year}{2003}): \emph{\bibinfo{title}{CellDesigner: a process diagram
  editor for gene-regulatory and biochemical networks}}.
\newblock {\sl \bibinfo{journal}{BIOSILICO}}
  \bibinfo{volume}{1}(\bibinfo{number}{5}), pp. \bibinfo{pages}{159--162},
  \doi{10.1016/S1478-5382(03)02370-9}.

\bibitemdeclare{article}{galaris_pantopoulos_2008}
\bibitem{galaris_pantopoulos_2008}
\bibinfo{author}{D.~\surnamestart Galaris\surnameend} \&
  \bibinfo{author}{K.~\surnamestart Pantopoulos\surnameend}
  (\bibinfo{year}{2008}): \emph{\bibinfo{title}{{O}xidative stress and iron
  homeostasis: mechanistic and health aspects}}.
\newblock {\sl \bibinfo{journal}{Crit Rev Clin Lab Sci}}
  \bibinfo{volume}{45}(\bibinfo{number}{1}), pp. \bibinfo{pages}{1--23},
  \doi{10.1080/10408360701713104}.

\bibitemdeclare{article}{granvilliers_benhamou_2006}
\bibitem{granvilliers_benhamou_2006}
\bibinfo{author}{L.~\surnamestart Granvilliers\surnameend} \&
  \bibinfo{author}{F.~\surnamestart Benhamou\surnameend}
  (\bibinfo{year}{2006}): \emph{\bibinfo{title}{{Algorithm 852}: {RealPaver}:
  an interval solver using constraint satisfaction techniques}}.
\newblock {\sl \bibinfo{journal}{{ACM} Transactions on Mathematical Software}}
  \bibinfo{volume}{32}(\bibinfo{number}{1}), pp. \bibinfo{pages}{138--156},
  \doi{10.1145/1132973.1132980}.

\bibitemdeclare{article}{hunt_marshall_1986}
\bibitem{hunt_marshall_1986}
\bibinfo{author}{R.~C. \surnamestart Hunt\surnameend} \&
  \bibinfo{author}{L.~\surnamestart Marshall-Carlson\surnameend}
  (\bibinfo{year}{1986}): \emph{\bibinfo{title}{{I}nternalization and recycling
  of transferrin and its receptor. {E}ffect of trifluoperazine on recycling in
  human erythroleukemic cells}}.
\newblock {\sl \bibinfo{journal}{J. Biol. Chem.}}
  \bibinfo{volume}{261}(\bibinfo{number}{8}), pp. \bibinfo{pages}{3681--3686}.

\bibitemdeclare{article}{knutson_oukka_2005}
\bibitem{knutson_oukka_2005}
\bibinfo{author}{M.~D. \surnamestart Knutson\surnameend},
  \bibinfo{author}{M.~\surnamestart Oukka\surnameend}, \bibinfo{author}{L.~M.
  \surnamestart Koss\surnameend}, \bibinfo{author}{F.~\surnamestart
  Aydemir\surnameend} \& \bibinfo{author}{M.~\surnamestart
  Wessling-Resnick\surnameend} (\bibinfo{year}{2005}):
  \emph{\bibinfo{title}{{I}ron release from macrophages after
  erythrophagocytosis is up-regulated by ferroportin 1 overexpression and
  down-regulated by hepcidin}}.
\newblock {\sl \bibinfo{journal}{Proc. Natl. Acad. Sci. U.S.A.}}
  \bibinfo{volume}{102}(\bibinfo{number}{5}), pp. \bibinfo{pages}{1324--1328},
  \doi{10.1073/pnas.0409409102}.

\bibitemdeclare{}{maxima}
\bibitem{maxima}
\bibinfo{author}{\surnamestart Maxima\surnameend} (\bibinfo{year}{2011}):
  \emph{\bibinfo{title}{Maxima, a Computer Algebra System. Version 5.24.0}}.
\newblock \urlprefix\url{http://maxima.sourceforge.net/}.

\bibitemdeclare{article}{mayr_janecke_2010}
\bibitem{mayr_janecke_2010}
\bibinfo{author}{R.~\surnamestart Mayr\surnameend}, \bibinfo{author}{A.~R.
  \surnamestart Janecke\surnameend}, \bibinfo{author}{.~\surnamestart
  Schranz\surnameend}, \bibinfo{author}{W.~J.H. \surnamestart
  Griffiths\surnameend}, \bibinfo{author}{W.~\surnamestart Vogel\surnameend},
  \bibinfo{author}{A.~\surnamestart Pietrangelo\surnameend} \&
  \bibinfo{author}{H.~\surnamestart Zoller\surnameend} (\bibinfo{year}{2010}):
  \emph{\bibinfo{title}{Ferroportin disease: A systematic meta-analysis of
  clinical and molecular findings}}.
\newblock {\sl \bibinfo{journal}{Journal of Hepatology}}
  \bibinfo{volume}{53}(\bibinfo{number}{5}), pp. \bibinfo{pages}{941--949},
  \doi{10.1016/j.jhep.2010.05.016}.

\bibitemdeclare{article}{moirand_mortaji_1997}
\bibitem{moirand_mortaji_1997}
\bibinfo{author}{R.~\surnamestart Moirand\surnameend}, \bibinfo{author}{A.~M.
  \surnamestart Mortaji\surnameend}, \bibinfo{author}{O.~\surnamestart
  Lor{\'e}al\surnameend}, \bibinfo{author}{F.~\surnamestart
  Paillard\surnameend}, \bibinfo{author}{P.~\surnamestart Brissot\surnameend}
  \& \bibinfo{author}{Y.~\surnamestart Deugnier\surnameend}
  (\bibinfo{year}{1997}): \emph{\bibinfo{title}{A new syndrome of liver iron
  overload with normal transferrin saturation}}.
\newblock {\sl \bibinfo{journal}{The Lancet}}
  \bibinfo{volume}{349}(\bibinfo{number}{9045}), pp. \bibinfo{pages}{95--97},
  \doi{10.1016/S0140-6736(96)06034-5}.

\bibitemdeclare{article}{omholt_kefang_1998}
\bibitem{omholt_kefang_1998}
\bibinfo{author}{S.~W. \surnamestart Omholt\surnameend},
  \bibinfo{author}{X.~\surnamestart Kefang\surnameend},
  \bibinfo{author}{{\O}.~\surnamestart Andersen\surnameend} \&
  \bibinfo{author}{E.~\surnamestart Plahte\surnameend} (\bibinfo{year}{1998}):
  \emph{\bibinfo{title}{Description and Analysis of Switchlike Regulatory
  Networks Exemplified by a Model of Cellular Iron Homeostasis}}.
\newblock {\sl \bibinfo{journal}{Journal of Theoretical Biology}}
  \bibinfo{volume}{195}(\bibinfo{number}{3}), pp. \bibinfo{pages}{339--350},
  \doi{10.1006/jtbi.1998.0800}.

\bibitemdeclare{article}{shvartsman_fibach_2010}
\bibitem{shvartsman_fibach_2010}
\bibinfo{author}{M.~\surnamestart Shvartsman\surnameend},
  \bibinfo{author}{E.~\surnamestart Fibach\surnameend} \&
  \bibinfo{author}{Z.~I. \surnamestart Cabantchik\surnameend}
  (\bibinfo{year}{2010}): \emph{\bibinfo{title}{{T}ransferrin-iron routing to
  the cytosol and mitochondria as studied by live and real-time fluorescence}}.
\newblock {\sl \bibinfo{journal}{Biochem. J.}}
  \bibinfo{volume}{429}(\bibinfo{number}{1}), pp. \bibinfo{pages}{185--193},
  \doi{10.1042/BJ20100213}.

\bibitemdeclare{article}{wang_pantopoulos_2011}
\bibitem{wang_pantopoulos_2011}
\bibinfo{author}{J.~\surnamestart Wang\surnameend} \&
  \bibinfo{author}{K.~\surnamestart Pantopoulos\surnameend}
  (\bibinfo{year}{2011}): \emph{\bibinfo{title}{{R}egulation of cellular iron
  metabolism}}.
\newblock {\sl \bibinfo{journal}{Biochem. J.}}
  \bibinfo{volume}{434}(\bibinfo{number}{3}), pp. \bibinfo{pages}{365--381},
  \doi{10.1042/BJ20101825}.

\bibitemdeclare{article}{weissman_klausner_1986}
\bibitem{weissman_klausner_1986}
\bibinfo{author}{A.~M. \surnamestart Weissman\surnameend},
  \bibinfo{author}{R.~D. \surnamestart Klausner\surnameend},
  \bibinfo{author}{K.~\surnamestart Rao\surnameend} \& \bibinfo{author}{J.~B.
  \surnamestart Harford\surnameend} (\bibinfo{year}{1986}):
  \emph{\bibinfo{title}{Exposure of K562 cells to anti-receptor monoclonal
  antibody OKT9 results in rapid redistribution and enhanced degradation of the
  transferrin receptor.}}
\newblock {\sl \bibinfo{journal}{The Journal of Cell Biology}}
  \bibinfo{volume}{102}(\bibinfo{number}{3}), pp. \bibinfo{pages}{951--958},
  \doi{10.1083/jcb.102.3.951}.

\end{thebibliography}
\end{document}